\documentclass[superscriptaddress,twocolumn,amsmath,amssymb,aps,pre,english]{revtex4-2}

\usepackage{amsmath,amssymb,graphicx,xcolor}

\usepackage[unicode=true]{hyperref}
\begin{document}

\title{Speed-quality tradeoff in a 
physical 
model of molecular sorting
}

\author{Elisa Floris}
\affiliation{Department of Biology, University of Graz, Universitätsplatz 2, 8010 Graz, Austria}
\affiliation{Institute of Condensed Matter Physics and Complex Systems,
Department of Applied Science and Technology, Politecnico di Torino,
Corso Duca degli Abruzzi 24, 10129 Torino, Italy}
\author{Damiano Andreghetti}
\affiliation{Institute of Condensed Matter Physics and Complex Systems,
Department of Applied Science and Technology, Politecnico di Torino,
Corso Duca degli Abruzzi 24, 10129 Torino, Italy}
\affiliation{Istituto Nazionale di Fisica Nucleare (INFN), Italy}
\author{Thibaut Caplet}
\affiliation{Institute of Condensed Matter Physics and Complex Systems,
Department of Applied Science and Technology, Politecnico di Torino,
Corso Duca degli Abruzzi 24, 10129 Torino, Italy}
\affiliation{Istituto Nazionale di Fisica Nucleare (INFN), Italy}
\affiliation{Université Paris-Saclay, ENS Paris-Saclay, DER de Physique, 91190, Gif-sur-Yvette, France}
\author{Luca Dall'Asta}
\affiliation{Institute of Condensed Matter Physics and Complex Systems,
Department of Applied Science and Technology, Politecnico di Torino,
Corso Duca degli Abruzzi 24, 10129 Torino, Italy}
\affiliation{Istituto Nazionale di Fisica Nucleare (INFN), Italy}
\affiliation{Italian Institute for Genomic Medicine (IIGM) and Candiolo Cancer Institute IRCCS, str.~prov.~142, km 3.95, Candiolo (TO) 10060, Italy}
\author{Andrea Gamba}
\email{andrea.gamba@polito.it}
\affiliation{Institute of Condensed Matter Physics and Complex Systems,
Department of Applied Science and Technology, Politecnico di Torino,
Corso Duca degli Abruzzi 24, 10129 Torino, Italy}
\affiliation{Istituto Nazionale di Fisica Nucleare (INFN), Italy}
\affiliation{Italian Institute for Genomic Medicine (IIGM) and Candiolo Cancer Institute IRCCS, str.~prov.~142, km 3.95, Candiolo (TO) 10060, Italy}

\begin{abstract}
  Eukaryotic cells rely on membrane-mediated processes to compartmentalize biomolecules, counteracting diffusion-driven homogenization. These processes involve the selective sorting and packing of molecules into lipid vesicles, which are then dispatched to appropriate intracellular destinations. Previous works introduced an abstract statistical physics framework for studying this molecular distillation process, where the membrane was treated as static 
  and the focus was on  molecular aggregation and  extraction. Here, we extend this framework to explicitly incorporate dynamic membrane behavior, including bending, curvature generation, and changes in membrane size due to vesicle fusion and fission events. Sorting domains drive membrane curvature, leading to vesicle formation, detachment, and a molecular distillation process that alters membrane size.
Using mesoscopic modeling and numerical tools, we investigate the 
resulting interplay between membrane mechanics and molecular sorting. We determine a well-defined parameter region where vesicle fission and efficient molecular sorting occur, controlled by 
membrane rigidity, spontaneous curvature, and pressure difference across the membrane. We further identify a trade-off between speed and quality of molecular distillation: parameters that accelerate vesicle formation tend to reduce the quality of distillation, and vice versa.
We 
propose 
maximization of
the rate of negative entropy production 
as a natural criterion
to optimally balance these competing effects.
Simulations show that this quantity exhibits a 
distinct
maximum at intermediate values of spontaneous curvature, pressure, and membrane rigidity, demonstrating the viability of the criterion. In this context, optimal parameters 
emerge
naturally
from the coupled dynamics of molecular aggregation and membrane mechanics, suggesting a possible strategy for cellular sorting systems to strike an efficient balance between rapid vesicle formation and high sorting quality. 
\end{abstract}

\maketitle

\section{Introduction}

Within eukaryotic cells, a complex network of membrane-mediated processes effectively compartmentalizes biomolecules, counteracting the homogenizing effects of diffusion~\cite{MN08}.
Examples include  
clathrin-mediated endocytosis~\cite{Roux18}, the recycling or degradation of membrane molecules~\cite{CS18}, and the sorting of molecules derived from the Golgi apparatus~\cite{AB99} or the endoplasmic reticulum~\cite{ZPS+12}. 
In these processes, molecules are selectively sorted and packed into lipid vesicles, which are subsequently 
dispatched along the appropriate pathways.
Vesicle formation begins with biomolecules binding to the lipid membrane,
where they diffuse laterally due to the bilayer’s fluidity. Various direct and indirect interactions drive their aggregation, leading to domain formation. These domains recruit molecules that promote membrane bending and fission, ultimately resulting in vesicle detachment. 
Abstracting from
the molecular complexity of vesicle formation, 
one can formulate
a minimal statistical-physics model 
capturing the main mesoscopic features of the process
~\cite{ZDG19,ZVS+21,FPP+22,PFD+23}.
In this framework, molecular
sorting can be viewed as  
a distillation process driven by phase separation, where
molecules concentrate in localized sorting domains,
and are eventually removed
from the membrane and carried away
into enriched vesicles. In the steady state, the system is characterized by three key parameters: the molecular influx \( \phi \), the average surface density of molecules \( \rho \), and the mean residence time of a molecule \(\bar T \).   These quantities are constrained by the steady-state relation~$\rho=\bar T \phi$~\cite{ZDG19}. This relation allows 
the efficiency of the process, measured in terms of the sorting rate $\bar{T}^{-1}$,
to be determined indirectly 
from the density of molecules on the membrane. 
As shown in Ref.~\cite{ZVS+21},
the optimal efficiency is attained
for intermediate strengths of the interaction 
that drives molecular aggregation,
a result that aligns with experimental findings where molecular attraction can be directly controlled
and higher sorting efficiency is similarly observed at intermediate interaction strengths
~\cite{DKW+21}.
Intrinsic to the phase-separation paradigm is the concept of
a critical size for sorting domains: domains below this size shrink and dissolve (``unproductive'' domains), while those above it continue growing until they are packed into a lipid vesicle (``productive'' domains),
a feature again consistent with main experimental findings
~\cite{FPP+22,WCM+21}. 

In Ref.~\cite{ZVS+21}, the membrane was considered as static, focusing on the attachment and detachment of molecules during vesicle formation. In the present study, the model is extended to explicitly account for membrane dynamics, including bending, curvature induction, and changes in membrane size due to fusion and 
fission events. The formation of sorting domains induces membrane curvature, as supported by experimental studies~\cite{SHS10,SSR+12,BHH+15,SHG+17,YAB+21,DKW+21}.
Membrane curvature can be generated through various mechanisms, such as asymmetries in lipid composition, conical-shaped transmembrane molecules, the insertion of molecules with hydrophobic motifs into the bilayer, or scaffolding by molecular domains and the cytoskeleton~\cite{MB15}. 
Protein
crowding, steric repulsions, and phase separation of intrinsically disordered proteins have also been reported to drive
membrane bending and fission~\cite{SHS10,SSR+12,BHH+15,SHG+17,YAB+21,DKW+21,WLM+24}.
The 
localization of curvature in sorting domains
can drive the ultimate
 detachment of membrane portions, creating a dynamic process where membrane regions are inserted and extracted. As a result, the size of the membrane changes dynamically through a distillation-like process driven by domain formation~\cite{ZVS+21}.

To achieve a deeper understanding of the molecular sorting phenomenon, we investigate these mechanistic aspects on a mesoscopic scale using phenomenological arguments and numerical tools. 
We find that efficient sorting, in the context of a dynamically changing membrane model,
requires specific combinations of membrane rigidity, spontaneous curvature, and pressure difference across the membrane.
These parameters control the occurrence
and timescale
of fission events.
By systematically varying them, we identify a fundamental trade-off between the speed of vesicle formation and the quality of molecular distillation: parameters that accelerate fission events (such as high spontaneous curvature or high membrane rigidity) lead to faster sorting, but produce vesicles with mixed molecular populations, whereas parameters that slow down the process (such as high pressure) improve sorting fidelity at the cost of reduced throughput.

\section{Membrane-molecule coupling}

The general, abstract scenario considered is that of a larger compartment, such as a cisterna of the Golgi apparatus, a part of the endoplasmic reticulum, or an endosome, that receives an influx of small vesicles carrying a mixture of molecules of a number of distinct species.
Although the process could be easily generalized to include a larger number of species, here we  focus on the case of two species only, 
which  will be conventionally
referred to  as~A~and~B.
Both species diffuse laterally on the compartment's membrane and aggregate into
homogeneous domains,
primarily
enriched with either A~or~B, driven by homotypic mutual affinity. The formation of these domains induces membrane invagination, ultimately leading to the 
fission of small vesicles containing the sorted molecules. 
\begin{figure}[!t]
\includegraphics[width=\linewidth]{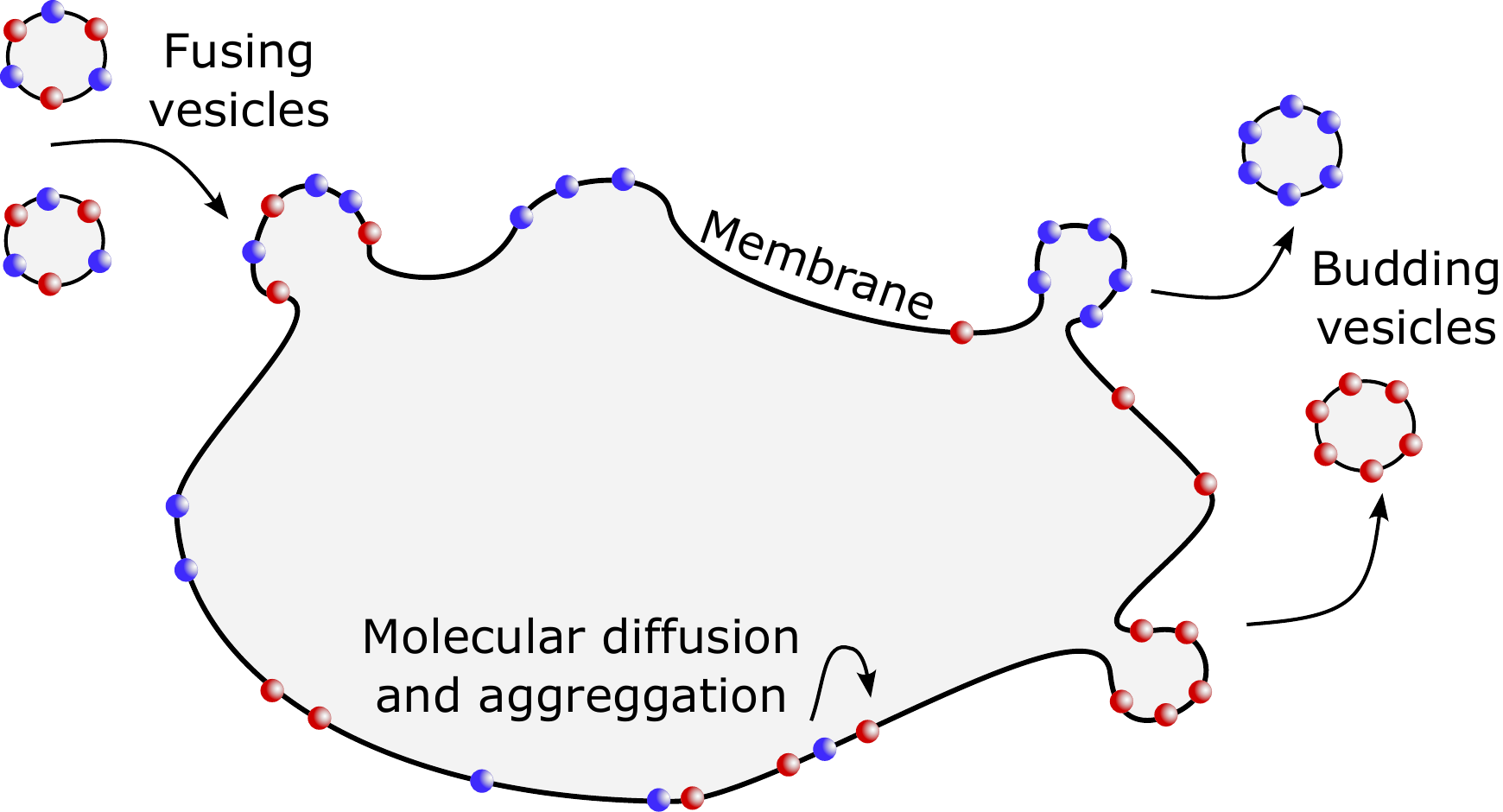}
    \caption{Schematic representation of the molecular sorting process. A~large lipid compartment is continuously supplied with vesicles containing a mixture of molecular species, here depicted as red and blue dots. Molecules diffuse laterally on the membrane and aggregate into localized domains due to specific chemical affinities. These domains induce local membrane bending, which eventually leads to the detachment of vesicles enriched in single molecular species.
 }
    \label{fig:schemino}
\end{figure}
As vesicles continuously arrive 
carrying their mixed-species cargo, 
the membrane acts as a dynamic sorting platform. The two species, initially intermixed, segregate into distinct domains, which in turn drive membrane deformation and vesicle budding. This process results in the selective extraction of vesicles enriched with only one of the two species (Fig.~\ref{fig:schemino}).

The process is studied at a mesoscopic scale, where the lipid bilayer of the membrane compartment is treated as a
closed surface~$\mathcal{S}$, with area $A[\mathcal{S}]$ and enclosing a volume $V[\mathcal{S}]$. At this scale,  
the formation of molecular domains 
influences membrane dynamics
by inducing a local shape deformation.
To describe the lipid compartment and the molecules diffusing on it, we consider the free-energy functional
\begin{equation}
    \label{eq:G}
    G[\mathcal{S}, \rho,\psi] = G_\mathrm{mem}[\mathcal{S}, \rho,\psi] + G_{\mathrm{int}}[\rho,\psi],
\end{equation}
where the first term couples membrane bending energy with total molecule density $\rho$ and composition parameter $\psi$. The latter tends to $1$ (resp. $-1$) within a domain of $A$ (resp. $B$) molecules and approximately $0$ when the membrane is not locally enriched in one of the two species. The second term of Eq.~\ref{eq:G}, $G_\mathrm{int}$, is the Flory-Huggins free-energy for two solute species with homotypic interactions.
The membrane term reads
\begin{equation}
    \begin{aligned}
    G_\mathrm{mem}[\mathcal{S}, \rho,\psi] &= \int_\mathcal{S} \mathrm{d}\mathbf{r} \,\frac{\kappa_m}{2}\, \left[H(\mathbf{r}) - H_0(\rho(\mathbf{r}),\psi(\mathbf{r}))\right]^2\\
    &\phantom{=}+ \sigma_m A[\mathcal{S}] - p_m\, V[\mathcal{S}],
    \end{aligned}
\end{equation}
where $\kappa_m$ is
the bending rigidity of the membrane, $H$ is the sum of local curvatures, and $H_0$ is 
the
local spontaneous curvature induced by the presence of molecules of the same species. Then there is a surface tension term, and the last term takes into account the effect of pressure difference $p_m>0$ between inner and outer regions of the membrane. The tension and pressure terms can also act as Lagrange multipliers for fixed area or volume ensembles. Systems described by similar Hamiltonians have been studied previously,
but not as a model for the non-equilibrium molecular sorting process considered here~\cite{Far11,Dur22,FMR+94}.

To obtain a tractable model, we integrate out the Gaussian fluctuations of the total density around its mean, $\rho(\mathbf{r}) = \rho_0 + \delta\rho(\mathbf{r})$. 
Assuming a homogeneous mean density $\rho_0$ is justified by our focus on steady-state properties of this system, where as a first approximation the total density can be considered nearly constant.
Near the demixing transition, we expand both the Flory-Huggins free energy and elastic energy terms for small composition ($\psi \ll 1$). Truncating to the lowest significant orders leads to the effective phenomenological functional
\begin{equation}
    \begin{aligned}
    \label{eq:G_GL}
    G_\mathrm{ph}[\mathcal{S}, \psi]&=\int_\mathcal{S} \mathrm{d}\mathbf{r}\left\{ 
    \frac{D}{2}|\nabla_\mathcal{S} \psi|^2+\frac{r}{2}\psi^2+\frac{u}{4}\psi^4\right\}\\
    &+\int_\mathcal{S} \mathrm{d}\mathbf{r}\left\{ \frac{\kappa_m}{2}H^2-\kappa c_0H\psi^2
    \right\}\\
    &+\sigma_m \,A[\mathcal{S}]-p_m\,V[\mathcal{S}],
    \end{aligned}
\end{equation}
where $D$ is the diffusivity, $\nabla_\mathcal{S}$ is the surface gradient, $r<0$ and $u>0$ are phenomenological parameters controlling the shape of the double-well potential inducing phase separation, and $c_0$ is the parameter controlling how intense is the spontaneous curvature induced locally by the presence of molecules of the same species. 

In the present setting, vesicles attach to the membrane and molecules contained within them
undergo diffusive dynamics on the membrane, with direct interactions driving the aggregation of molecules of the same species. These interactions 
represent  chemical affinities between molecules, leading to the formation of domains that induce a perturbation of membrane shape. Depending on domain size and membrane parameters the membrane engulfment can eventually lead to the formation of a vesicle detaching from the original  compartment.
\begin{figure*}[t]
    \centering    \includegraphics[width=\textwidth]
{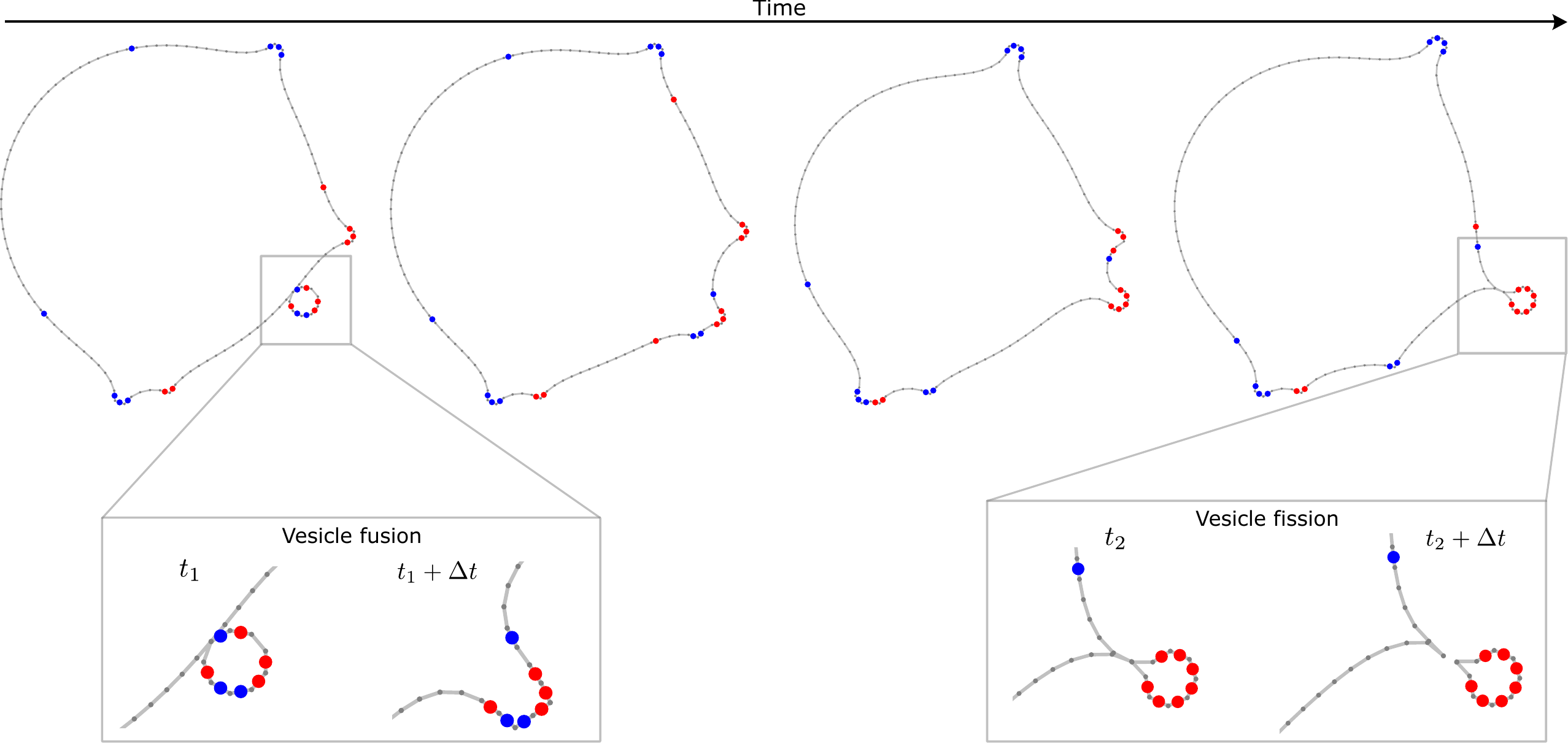}
    \caption{Time evolution of a representative membrane configuration obtained 
    from numerical simulations.
Molecular dynamics drives aggregation, while membrane relaxation leads to the engulfment of newly formed aggregates. In the box on the left a vesicle fusion event is shown, consisting in the attachment of a vesicle to the membrane compartment. The fusing vesicle carries a mixture of A and B molecules (marked in red and blue, respectively). In the box on the right a vesicle fission event is shown. The formation of a vesicle, driven by membrane bending, is completed when the membrane undergoes self-intersection. The vesicle is then detached from the larger membrane compartment.
}
    \label{fig:example}
\end{figure*}
Vesicle fusion and fission are highly complex biophysical processes involving the interaction of a plethora of molecular factors.
Here, we abstract  from these  complicated molecular details, and simply assume that 
fusion and fission take place instantly when  an incoming vesicle makes contact with the membrane system, or a closed 
vesicle, ready to be detached, is formed from the host membrane.  Even in this simplified form, 
fusion and fission events introduce significant complexity, primarily because they entail topological changes in the simulated membrane, which pose a major challenge for mesoscale simulations.

\section{Numerical model}

While vesicle fission and fusion have mainly been studied through particle-based molecular dynamics simulations, focusing on molecular-level details~\cite{LLW+09,HSC+21}, we aim to investigate their impact on larger scales, where they influence sorting efficiency.
 These processes involve the continuous creation and removal of membrane patches and self-intersection events, which standard numerical schemes typically avoid.

As a first step in the study of this problem, we simulate here a one-dimensional system, where the membrane is represented as a closed elastic curve. This approach allows us to capture key aspects of the dynamics while avoiding the computational difficulties associated with handling membrane self-intersections in higher dimensions. We 
therefore restrict our analysis to 
qualitative effects that are expected to be independent of dimensionality, such as the interplay between membrane elasticity and sorting dynamics. 
As demonstrated in previous studies~\cite{FB08,Far11,Dur22}, qualitative insights obtained from one-dimensional models often extend to higher-dimensional cases. 

To numerically simulate the process, we assume that the membrane relaxes to its equilibrium shape through a Langevin dynamics. Molecular dynamics is discretized using a lattice-gas kinetic scheme~\cite{ZVS+21,FPP+22,PFD+23}.
Molecules undergo diffusive dynamics on the membrane, with direct interactions driving the aggregation of molecules of the same species. 

We discretize the membrane surface into a one-dimensional periodic lattice of $N$ nodes in the plane, connected by oriented edges 
of fixed length~$a$. The oriented edges are represented as planar vectors~$\mathbf{y}_k$, with $k \in \{1,2,...,N \}$. The presence or absence of a molecule on a given edge is described by an occupation number $\sigma_k \in \{0, \pm 1\}$, where $\pm 1$ corresponds to occupation by a molecule of the~A or~B species, respectively,
and~$0$ corresponds to an unoccupied site. 
The energy for the discretized system is

\begin{equation}
    H=H_{\mathrm{mem}}+H_{\mathrm{int}},
    \label{eq:discr_ham}
\end{equation}
where
\begin{align}
\nonumber H_{\mathrm{mem}} & =\frac{\kappa}{2} \sum_{k=1}^N (c_k-c_{0,k})^2 a  - p A  
\\& \quad  + \mathbf{h} \cdot \sum_{k=1}^N \mathbf{y}_k +\sum_{k=1}^N \lambda_k (|\mathbf{y}_k|-a),
\label{eq:discr_mem_ham}
\end{align}
and
\begin{equation}
    H_{\mathrm{int}}=-k_BT\log g \sum_{\langle i,j\rangle} (\delta_{\sigma_i,\sigma_j}-\delta_{\sigma_i,0}\delta_{\sigma_j,0}).
    \label{eq:discr_mol_ham}
\end{equation}

In the discretized version of the membrane Hamiltonian $H_{\mathrm{mem}}$, the total area enclosed by the membrane is denoted by $A$, the curvature at each vertex by $c_k$, and the spontaneous curvature by $c_{0,k}$ that depends on the colocalization of same-type molecules in the same region (see App.~\ref{app:numerical_scheme}). The Lagrange multipliers $\mathbf{h}$ and $\lambda_k$ enforce membrane closure and contour length constraints, respectively. The Hamiltonian of molecular inclusions $H_{\mathrm{int}}$ describes a one-dimensional multiple-species lattice-gas, with a non-dimensional interaction constant $g$.
Membrane relaxation is implemented through Langevin dynamics in the zero-temperature limit, reducing the dynamics to a deterministic gradient descent on the energy landscape. This allows us to study the interplay between aggregation and bending, while retaining a framework that can be straightforwardly extended to finite temperature in future work.
New molecules are introduced into the system through vesicle fusion events taking place at a rate~$k_I$. 
Molecules diffuse and aggregate on the membrane 
according to a Kawasaki dynamics~\cite{kaw66}:
the occupation number $\sigma_k$ is exchanged with its nearest neighbor $\sigma_{k+1}$ with rate
\begin{equation}
r_D=k_D e^{-\beta\Delta H_{\mathrm{int}}},
\end{equation}
where $k_D$ is a diffusion rate, $\beta=1/k_BT$, and $\Delta H_{\mathrm{int}}$ is the energy difference associated with the exchange.

This choice of exchange rate implicitly assumes that the short-range molecular interaction controlled by $g$ is much stronger than the curvature-mediated coupling, whose leading effect on individual hops then reduces to a renormalization of the effective molecular diffusivity. Moreover, since $c_{0,k}$ is only non-zero for a pair of like neighbors, membrane curvature is a genuinely collective effect requiring at least two co-localized molecules, rather than one generated by single-particle hops.
The formation of molecular domains is thus driven purely by the direct interactions between molecules entering $H_\mathrm{int}$, with strength controlled by the dimensionless parameter~$g$. On a coarse-grained scale, this microscopic Kawasaki kinetics maps onto the quartic interaction term of the phenomenological functional $G_\mathrm{ph}$ in Eq.~\eqref{eq:G_GL}, with the composition field $\psi$ playing the role of the coarse-grained occupation $\sigma_k$. 

Whenever membrane self-intersection events occur, the enclosed segment of the membrane is removed, simulating a fission event.
Fusion and fission events affect membrane size $N$, as lipids are moved in and out of the compartment. 
We study the system at the steady state,
where the compartment size fluctuates around its average value.  
Unless otherwise specified, all simulations are performed with initial membrane size $N=100$, $\beta \kappa = 40a$, $k_I/k_D=10^{-4}$ and $g=8$.
Additional details of the numerical scheme are provided in App.~\ref{app:numerical_scheme}.

In this framework, we simulated the time evolution of a membrane compartment subject to a steady influx of  vesicles
carrying a mixed population of molecules.
We observed that molecular aggregation on 
the membrane leads to the nucleation of sorting domains, which in turn drive progressive membrane engulfment. This sequence is illustrated in  Fig.~\ref{fig:example}, where consecutive snapshots show molecules of the same species forming aggregates that locally induce  membrane bending.

\section{Parametric conditions for molecular sorting}
\label{sec:steady-state-condition}

Molecular sorting requires continuous influx of mixed vesicles and distillation of purified ones.
The coupling of membrane elasticity with molecular concentrations imposes constraints that must be met for a vesicle to form.
If the extraction of vesicles from the sorting compartment is halted,
the stationary state cannot be reached and the system gets congested with molecules. 
Here we will discuss the energetic conditions for which
vesicle budding is favored, and will investigate numerically the parameter region where a steady state is reached. We start by considering the one-dimensional counterpart of Eq.~\eqref{eq:G_GL} corresponding to our numerical scheme
\begin{equation}
    \begin{aligned} \label{eq:1dcontinuous}
    G_\mathrm{ph}&=\int_\Omega \mathrm{d}s\left\{ 
    \frac{D}{2}(\partial_s\psi)^2+\frac{r}{2}\psi^2+\frac{u}{4}\psi^4\right\}\\
   & +\int_\Omega \mathrm{d}s\left\{\frac{\kappa}{2}H^2-\kappa c_0H\psi^2
    \right\}-p\,A[\Omega],
    \end{aligned}
\end{equation}
where now the membrane is represented by the curve~$\Omega$. The phenomenological parameters $D$, $r$, $u$ and $c_0$ entering Eq.~\eqref{eq:1dcontinuous} can be systematically obtained from the coarse-graining of $H_\mathrm{mem}$ and $H_\mathrm{int}$ in Eqs.~\eqref{eq:discr_mem_ham} and \eqref{eq:discr_mol_ham} via a mean-field treatment of the lattice-gas model, together with a gradient expansion of the composition field. The derivation, together with the explicit expressions relating the coarse-grained parameters to the microscopic ones, is given in App.~\ref{app:coarse_graining}.

When a domain of one species of molecules is formed, the interface profile takes the usual form~\cite{C77}
\begin{equation}
    \psi(s)\approx \psi_b \tanh \left( 
    \frac{s}{\xi}
    \right),
\end{equation}
where $\psi_b = \sqrt{(\kappa c_0 H-r)/u}$ is the composition parameter in the bulk phases and $\xi=\sqrt{D/(\kappa c_0 H -r)}$ is the interface width.  Notably, the presence of a coupling between molecules and membrane curvature introduces an additional contribution to the line tension~\cite{L92}. Indeed, integrating the energy over the profile yields the line tension
\begin{equation}
    \gamma =\frac{2\sqrt{2}}{3} \frac{ \sqrt{D}}{u}(\kappa c_0 H-r)^{3/2},
\end{equation}
where membrane curvature $H$ is assumed to be homogeneous inside the domain.

\begin{figure}
    \includegraphics[width=0.9\linewidth]{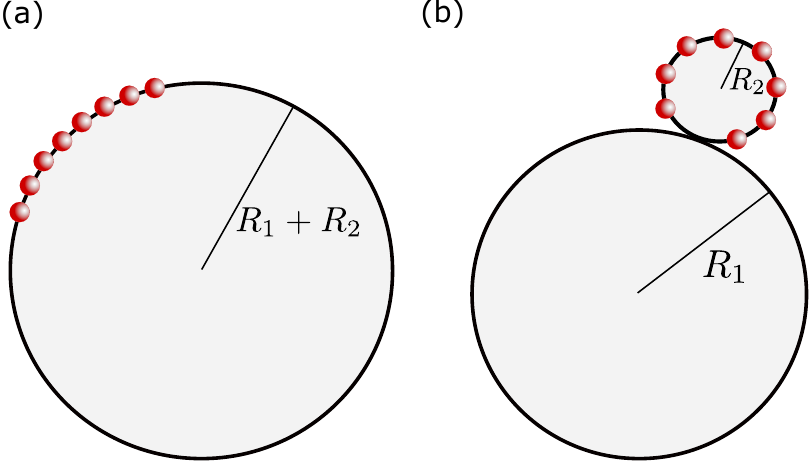}
    \caption{Schematic representations of (a) a configuration in which a homogeneous molecular domain has formed, but fission does not occur, and (b) a configuration in which fission has occurred.}
    \label{fig:fission_or_not}
\end{figure}

For vesicle budding to be energetically favorable, a balance between curvature and pressure
is required. The elastic energy term, in the presence of a molecular domain, induces membrane bending 
and vesicle budding
through coupling with the spontaneous curvature~$c_0$. However, since the area enclosed by the membrane would be reduced, the pressure term impairs the budding event.
This suggests that there 
exists a specific region in the $(c_0, p)$ parameter space where the sorting process can occur.

\begin{figure*}
    \centering
    \includegraphics[width=0.7\linewidth]
    {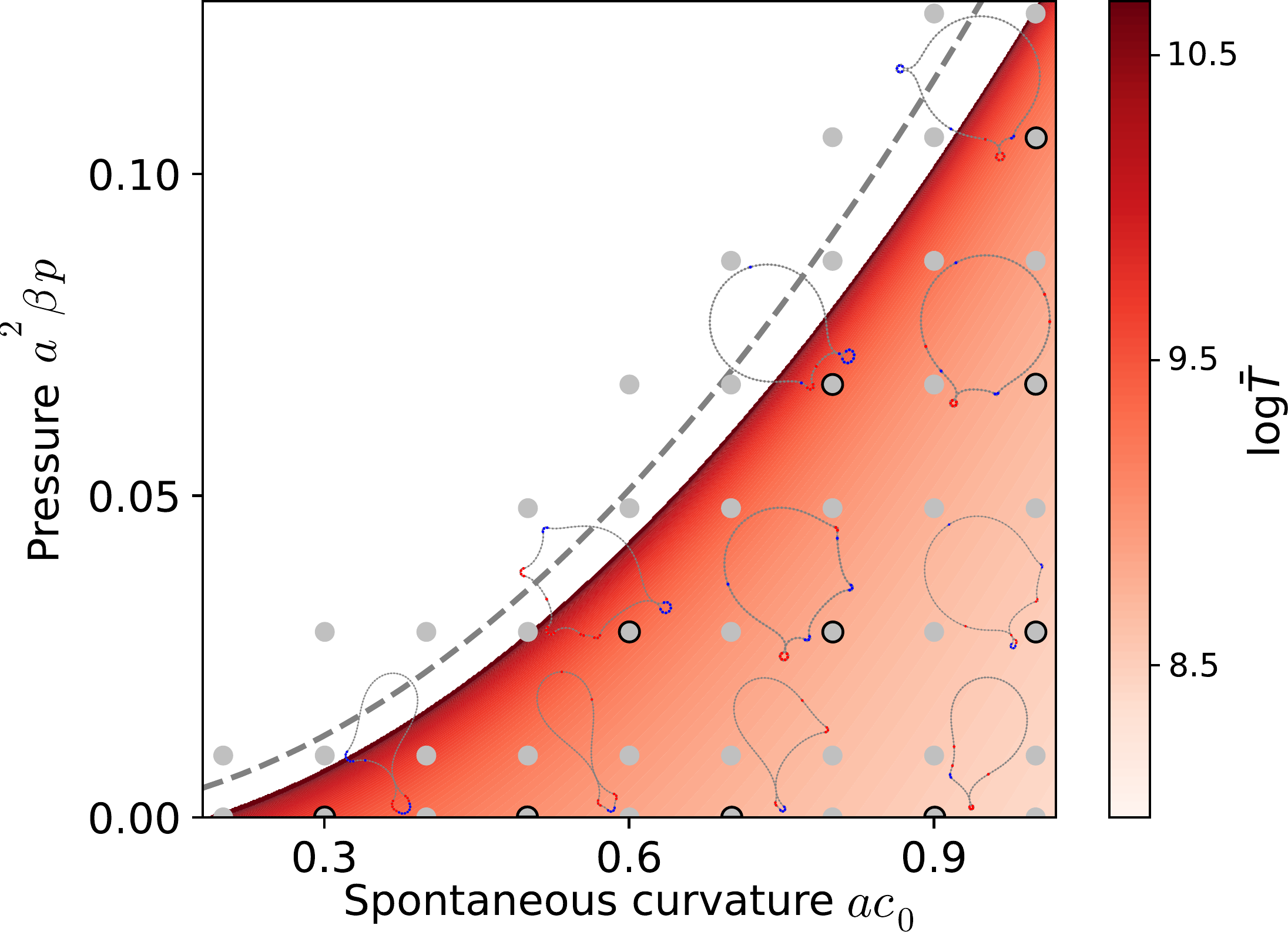}
    \caption{Heatmap of 
    the logarithmic
    mean residence time,
    $\log \bar T$, as a function of pressure $p$
    and spontaneous curvature
    $c_0$,
    at fixed membrane rigidity $\beta\kappa/a=40$ (dimensionless parameters are used in the graph). 
    In the upper-left region of
    this parameter space, 
    no vesicle fission occurs, preventing the membrane from reaching a stationary state. The
    approximately
    parabolic boundary 
    with the region where the system reaches a stationary state
    is consistent with the estimate Eq.~\ref{eq:pressure_condition}  (gray dashed line). 
    Gray dots correspond to values of $\log \bar T$
    directly computed from the simulations. 
    The remaining values are obtained by interpolation.
    Representative membrane configurations, corresponding to the black-circled dots, are shown. At high pressure values, membrane configurations 
    become more regular.}
    \label{fig:heatmap_cprot_pr}
\end{figure*}

To estimate this effect we compute the energy difference between two configurations: one where the vesicle is closed and one where the domain does not bend the membrane (see Fig.~\ref{fig:fission_or_not} and App.~\ref{app:budding} for calculations). One finds that in the limit $R_1\gg R_2$, and assuming $R_2\sim c_0^{-1}$, in order for the vesicle to close, the pressure should satisfy
\begin{equation}
\label{eq:pressure_condition}
 p<p^*=\frac{\gamma c_{0}}{\pi R_{1}}+\frac{\kappa c_{0}^{2}}{2R_{1}}+O\left(R_{1}^{-2}\right).    
\end{equation}

Thus, for the formation of a closed spherical vesicle, there exists a threshold value for the pressure, determined by the curvature imposed by the molecules and the line tension of the domain. This condition is analogous to the one derived for the 
three-dimensional case in previous studies \cite{Hel73,MFB+91,FMR+94}.

To more precisely determine  a phase diagram for molecular sorting 
we turn to numerical simulations, which fully capture the 
complex interplay of 
membrane shapes (beyond the spherical approximation considered for deriving Eq.~\ref{eq:pressure_condition})
and sorting dynamics. Fig.~\ref{fig:heatmap_cprot_pr} shows the mean residence time of molecules on the membrane, $\bar T$, measured for various values of 
the nondimensionalized pressure and curvature,
$a^2 \beta p$ and $a c_0$. The simulations confirm the qualitative picture gained from Eq.~(\ref{eq:pressure_condition}): for every value of~$c_0$, there exists a threshold $p^*(c_0)$ above which no fission occurs, the  membrane grows indefinitely, and no stationary state is reached. 
Fig.~\ref{fig:heatmap_cprot_pr} also provides intuition about the 
typical morphologies observed in different regions of the 
phase diagram.
No configuration is shown for the upper-left region, where no steady state is achieved. The figure provides visual evidence that an increase in pressure causes the membrane to adopt a more circular shape, while a greater spontaneous curvature favors invagination and promotes budding events.

\section{Speed-quality trade-off}
\label{sec:speedquality}

\begin{figure}
    \centering
\includegraphics[width=0.8\linewidth]{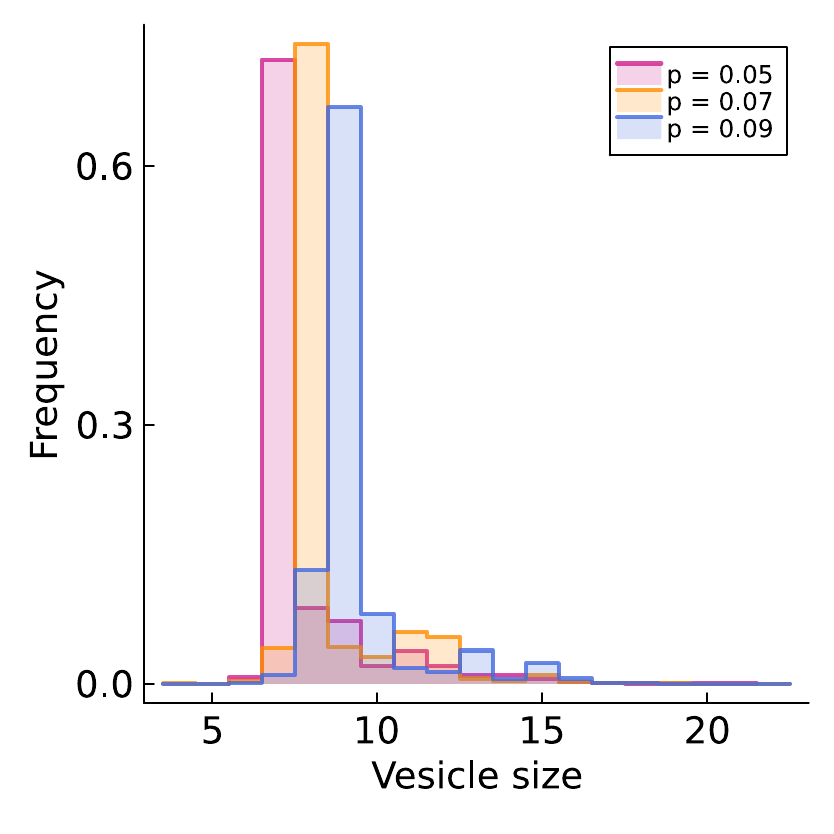}
    \caption{Normalized histogram of linear sizes of extracted vesicles (in lattice units $a$) for three different values of the (dimensionless) pressure. The size distributions are peaked at a characteristic value $ 2\pi R^*$ which is a function of membrane parameters. Simulations for these plots were run with spontaneous curvature $ac_0=0.9$ and membrane rigidity $\beta \kappa/a=40$.}
    \label{fig:extraction_size_histogram}
\end{figure}

In the present framework,
molecular sorting is viewed as an out-of-equilibrium coagulation process, where the rate of molecular distillation is determined by the frequency of fusion and fission events. 
In our previous work, fission was modeled through a fixed  rule, where  all domains exceeding 
an imposed phenomenological threshold
were instantly removed from the system~\cite{ZVS+21, FPP+22,
PFD+23}. 
Here, a characteristic
extraction size
instead emerges naturally from the interplay between domain formation, driven by attractive intermolecular interactions, and membrane 
bending,
induced by the formation of these same domains. 
This is reflected in the numerical measurement of the probability $P_\mathrm{ext}(R)$ that an extracted domain has size $R$, shown in Fig.~\ref{fig:extraction_size_histogram}, that is clearly peaked at a typical vesicle size $R^*$ of the order of $1/c_0$ and growing when increasing the pressure $p$, as predicted in Appendix~\ref{app:activation_barrier}.

This approach provides a more realistic representation of the endogenous mechanism underlying vesicle formation and sorting, revealing
how the physical properties of the membrane, such as its rigidity and 
the curvature imposed by molecular inclusions, can
regulate both the size of vesicles and the timescale of fission events. Rather than being externally imposed, a characteristic vesicle size emerges dynamically at steady state from the balance between membrane mechanics and the strength of intermolecular attraction. 
This equilibrium depends on competing factors: faster membrane engulfment accelerates vesicle distillation, but molecular domains require sufficient time to reorganize before fission occurs. If this rearrangement is too slow, vesicles will encapsulate mixed molecular populations, reducing the sorting efficiency. This interplay hints at the existence of
a fundamental trade-off between the speed and quality of the distillation process, raising the question of whether an optimal balance 
between these competing effects can be established.
To investigate this  
point, we define the quality of distillation as follows
\begin{equation}\label{eq:quality_def}
    q = \frac{|n_A-n_B|}{n_A+n_B},
\end{equation}
where $n_A$ and $n_B$ are the number of molecules of species $A$ and $B$, respectively, in the distilled vesicle. 
This gives
$0\leqslant q\leqslant 1$, with $q=0$ if both species are equally abundant and $q=1$ if the vesicle contains only one species.

\begin{figure*}[!ht]
\includegraphics[width=\linewidth]
{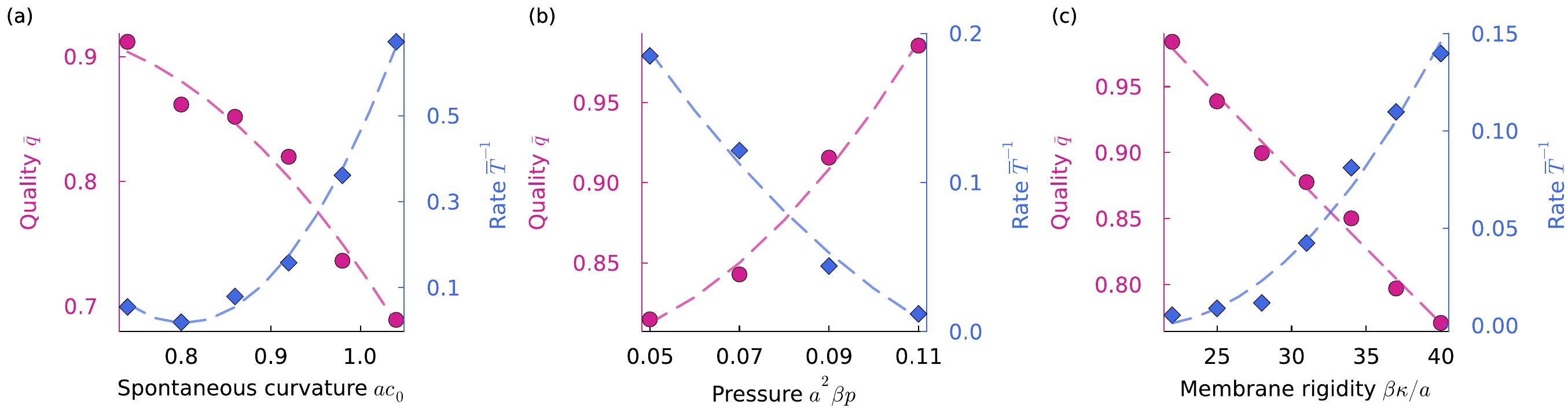}
    \caption{Measured mean
    sorting quality $\bar q$ (magenta dots) and sorting rate $\bar T^{-1}$ (blue diamonds) 
    as function of a single 
    control parameter, while keeping the others fixed
 (in the graphs, nondimensionalized parameters are used ). 
    (a)~Increasing spontaneous curvature 
$c_0$, 
    at fixed $\beta\kappa/a=40$ and $a^2\beta \, p=0.06$, facilitates vesicle formation, leading to a higher sorting rate, but lowers quality, as molecules are engulfed too rapidly to fully separate into homogeneous domains. No sorting process occurs for values  $a c_0<0.7$, as no stationary state is reached below this threshold. (b) Increasing pressure 
    $ p$,  
    at fixed $\beta\kappa/a=40$ and $a c_0=0.9$, slows down vesicle formation, allowing molecules to rearrange into demixed domains, and thus resulting in higher quality and lower sorting rate. (c) Increasing membrane rigidity 
    $\kappa$, 
    at fixed $a c_0=0.9$ and $a^2\beta  p=0.06$, promotes vesicle formation, but the resulting vesicles have 
    a more heterogeneous molecular composition. The dashed lines are
    parabolic fits, provided as a visual guide. Error bars
    are comparable in size to the symbols.
    } 
 \label{figure:tradeoff}
\end{figure*}
To study how the sorting quality~$\bar q$ and the 
rate~$\bar T^{-1}$
of the sorting process are influenced by each individual parameter, we systematically varied one parameter at a time: spontaneous curvature $c_0$, pressure $p$, or rigidity~$\kappa$, while keeping the others fixed. The
mean
sorting quality~$\bar q$ and sorting rate $\bar T^{-1}$ are shown in~Fig.~\ref{figure:tradeoff} as functions of the different parameters. The vesiculation process accelerates as both 
the spontaneous curvature  $c_0$ 
or membrane rigidity~$\kappa$ are increased~(Fig.~\ref{figure:tradeoff}(a,c)). This can be 
understood
as follows: increasing membrane rigidity enhances its responsiveness to induced deformations, facilitating faster engulfment of molecules into the developing vesicle. Similarly, increasing the spontaneous curvature~$c_0$ 
makes the membrane more prone to bending, thus promoting the distillation process. However, this increased speed comes at a cost: as the membrane encapsulates molecules more rapidly, there is less time for molecular rearrangement and the formation of homogeneous domains. As a result, the distillation quality is reduced, as the vesicles are less homogeneous and tend to contain a mixed molecular population. On the other hand, increasing the pressure $p$ has the opposite effect~(Fig.~\ref{figure:tradeoff}(b)). This parameter contributes to improve the sorting quality by slowing down the process: increased pressure promotes more controlled and gradual vesicle formation, allowing more time for molecular rearrangement into domains. 

For values of $p$ not too small, where vesicle shapes are regular and the size distribution is peaked, these trends can be understood analytically. Indeed, both the mean residence time 
$\bar{T}$
and the sorting quality are monotonically increasing functions of the typical extraction size $R^*$ (see App.~\ref{app:sorting_rate} and
\ref{app:quality} for details) and the latter grows with~$p$ and decreases with $c_0$ and $\kappa$
(see App.~\ref{app:activation_barrier}). A~larger typical vesicle size~$R^*$ (lower $c_0$ and/or $\kappa$, higher~$p$) gives a slower sorting rate~$\bar T^{-1}$ but a higher quality~$q$, and vice versa, so that the speed-quality trade-off is a direct consequence of how membrane mechanics sets the typical extracted vesicle size.

As discussed in the previous sections, for some parameter values, the system cannot sustain a steady state, and no effective sorting occurs. In Fig.~\ref{figure:tradeoff},
we only consider parameter sets that allow the system to reach a stationary state, 
yielding a well-defined sorting rate~$\bar T^{-1}$.

\section{Entropy production}

\begin{figure*}[!ht]
    \centering
    \includegraphics[width=0.9\linewidth]
    {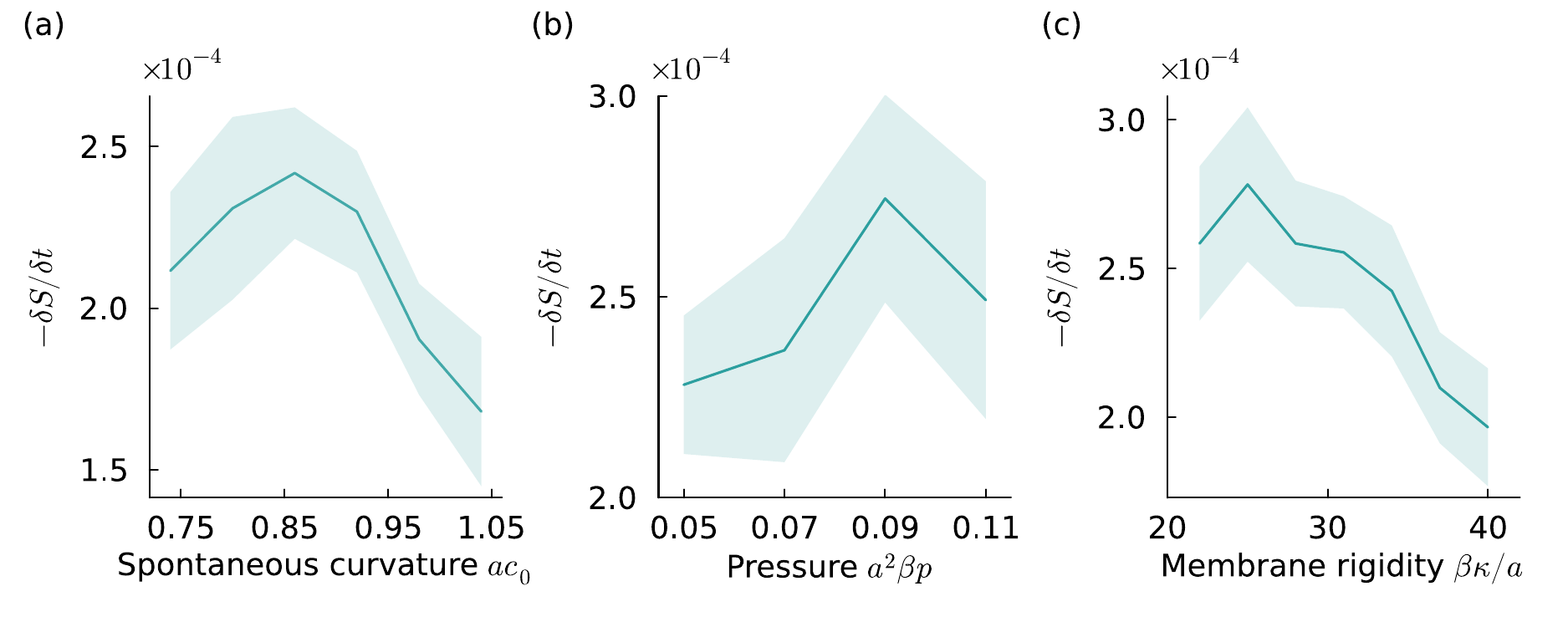}
    \caption{Rate of negative entropy production as a
function of (a) the spontaneous curvature
    $c_0$
at fixed membrane rigidity $\beta\kappa/a=40$
and pressure
$a^2\beta p=0.06$, (b) pressure
    $p$ 
at fixed rigidity $\beta\kappa/a=40$ and spontaneous curvature $a c_0=0.9$, (c)~membrane rigidity $\kappa$ at fixed spontaneous curvature $a c_0=0.9$ and pressure $a^2\beta p=0.06$. A clear maximum at intermediate values of the parameters is observed. Dimensionless parameters are used in the plots. Shaded areas: standard deviation.
    }
    \label{fig:entropy}
\end{figure*}

The speed-quality trade-off identified in the previous section raises a fundamental question: is there an optimal operating regime for the molecular sorting machinery? 
While sorting speed and quality impose opposing demands, biological systems likely evolved to balance these competing objectives. 

Here, we propose that the rate of negative entropy production provides a natural criterion for identifying such an optimum. Molecular sorting is a non-equilibrium process that actively creates order within the cell. Incoming vesicles carry high-entropy, well-mixed molecular populations, whereas outgoing vesicles have lower entropy due to their enrichment in specific species. The cell thus expends energy, through enzymatic processes, membrane bending, and fission, to reduce the entropy of the mixture. This ordering process can be quantified by the reduction of mixing entropy, a standard measure of disorder in a multi-component system~\cite{LL51,BSM+13}. 

We define the rate of negative entropy production as
\begin{eqnarray}
-\frac{\delta S}{\delta t} &=& -\frac{S_{\mathrm{out}} - S_{\mathrm{in}}}{\delta t},
\label{eq:negentropy}
\end{eqnarray}
where \( S_{\mathrm{in}} \) and \( S_{\mathrm{out}} \) are the
total
mixing entropies of the molecular populations residing on all the vesicles fusing with, or budding from, the membrane over a time interval~\( \delta t \), respectively. 
The mixing entropy contribution for each given vesicle containing \( n_\mathrm{A} \) and \( n_\mathrm{B} \) molecules of species A and B and $n_\mathrm{E}$ empty sites is

\begin{eqnarray}\label{eq:mixing_entropy}
s &=& -\rho_\mathrm{A} \log \rho_\mathrm{A} - \rho_\mathrm{B} \log \rho_\mathrm{B} - \rho_\mathrm{E} \log \rho_\mathrm{E},
\end{eqnarray}
with \( \rho_\mathrm{A} = n_\mathrm{A}/(n_\mathrm{A} + n_\mathrm{B} + n_\mathrm{E}) \), \( \rho_\mathrm{B} = n_\mathrm{B}/(n_\mathrm{A} + n_\mathrm{B} + n_\mathrm{E}) \) and \( \rho_\mathrm{E} = n_\mathrm{E}/(n_\mathrm{A} + n_\mathrm{B} + n_\mathrm{E}) \).

Writing $n_\mathrm{A}+n_\mathrm{B}=\rho(1\pm q)/2$ in terms of the total occupied fraction $\rho=\rho_\mathrm{A}+\rho_\mathrm{B}$ and the quality $q$ of Eq.~\eqref{eq:quality_def}, the mixing entropy, at fixed $\rho\approx \rho_0$, becomes, up to an additive constant, 

\begin{equation}
s(q) =  -\frac{\rho_0}{2}\big[(1+q)\log(1+q)+(1-q)\log(1-q)\big] ,
\label{eq:mixing_entropy_q}
\end{equation}
so that $\partial s/\partial q = -(\rho_0/2)\log[(1+q)/(1-q)]<0$ for all
$0< q< 1$:
the mixing entropy of a vesicle is a monotonically decreasing function of its sorting quality, and correspondingly, $\Delta s(q)=s(q_\mathrm{in})-s(q)$ is monotonically increasing in $q$.

The negative entropy production rate in Eq.~\eqref{eq:negentropy} depends on both the frequency of fission events and distillation quality;
 an~increase in either  of these  two competing
 factors
 contributes positively to the rate.

More precisely, weighting the per-vesicle entropy reduction $\Delta s(R)$ of a vesicle of size $R$ by the rate at which vesicles of that size are actually produced, 
namely
$J(R)N_\mathrm{st}(R)$ (with $J(R)$ the Kramers extraction rate and $N_\mathrm{st}(R)$ the stationary domain-size distribution of App.~\ref{app:budding} and Refs.~\cite{ZVS+21,FPP+22}), gives
\begin{equation}
-\frac{\delta S}{\delta t}\propto\int_{0}^\infty J(R)\,N_\mathrm{st}(R)\,\Delta s(R)\,\mathrm{d}R.
\label{eq:entropy_integral}
\end{equation}
In our simulations, the vesicle size distribution is sharply peaked around a typical value $R^*$ (see Fig.~\ref{fig:extraction_size_histogram} and App.~\ref{app:activation_barrier} for a theoretical interpretation), therefore  
Eq.~\eqref{eq:entropy_integral} is well approximated by $-\delta S/\delta t\approx \Delta s(q(R^*)) \Phi_v(R^*)$, where $\Phi_v(R^*) \approx N_\mathrm{st}(R^*)J(R^*)$ is the number of extracted vesicles of size $R^*$ per unit time. By mass balance, for domains of approximately constant internal density $\rho_d$, the extraction rate $\Phi_v(R^*)\approx \phi/(2\pi \rho_d R^*)$. On the other hand, the quality $q(R^*)$ grows with the vesicle size $R^*$
(see App.~\ref{app:quality}). The typical size~$R^*$ grows with $p$ and decreases with $c_0$ and~$\kappa$
(see App.~\ref{app:activation_barrier}), thus the extraction rate $\Phi_v(R^*)$ and the quality $q(R^*)$ respond in opposite directions to a change in each control parameter $c_0$, $p$, and~$\kappa$. Since $-\delta S/\delta t$ is proportional to a product of one factor that increases and one that decreases with the same parameter, its derivative with respect to any parameter will generically vanish at some intermediate value, 
corresponding to a maximum of the curve. 

This prediction is confirmed by numerical simulations, which show that an optimal balance can indeed be established. We computed \(-\delta S/\delta t\) across the same parameter ranges analyzed in Sec.~\ref{sec:speedquality}, finding clear maxima in the negative entropy production rate at intermediate values for each parameter (Fig.~\ref{fig:entropy}). In particular, when varying:\\
\textit{Spontaneous curvature} ($c_0$): 
At low \( c_0 \),
curvature is insufficient to drive efficient vesicle budding, leading to slow sorting. 
Conversely, at high \(  c_0 \), vesicles form so rapidly that molecules lack the time to segregate, resulting in poor distillation quality. An optimal curvature emerges at an intermediate value of $c_0$, where these effects are balanced (Fig.~\ref{fig:entropy}a).\\
\textit{Pressure}
($\Delta p$):  
Low pressure leads to irregular membrane shapes and inefficient
vesicle 
budding, while high pressure
tends to suppress fission.
The negative entropy production rate peaks at an intermediate pressure
value, where vesicle formation is frequent enough 
to sustain distillation, and slow enough to permit molecular organization (Fig.~\ref{fig:entropy}b).\\
\textit{Membrane rigidity} ($\kappa$): Low rigidity results in floppy membranes
that  fail to generate enough fission events. High rigidity 
on the other hand
resists curvature induction, slowing  down vesicle formation. An intermediate rigidity
value
 maximizes the rate of negative entropy production
 by providing the necessary 
 elastic response to drive budding without halting the process (Fig.~\ref{fig:entropy}c).

The presence of these maxima  supports the viability of the 
criterion: an optimal balance between speed and quality naturally emerges at intermediate parameter values
from the coupled dynamics of molecular aggregation and membrane mechanics.
From a biological perspective, it is tempting to speculate that living cells may have evolved to tune membrane properties, such as protein-induced curvature, membrane rigidity, and  
intracellular
pressure, to operate near maxima of negative entropy production. Such tuning would allow efficient conversion of metabolic work into the spatial order 
necessary
for cellular function.

\section{Conclusions}
The combined dynamics of lipid membranes and laterally diffusing particles, which depends on mechanical
properties such as membrane rigidity and 
curvature induced
by molecular inclusions, is highly complex, due to the nontrivial coupling of the two systems~\cite{GB+93,RIH+07,NAB09,Gov18}. 
In this work, we have moved beyond static approximations by introducing a mechanistic model that explicitly incorporates membrane bending, 
curvature induction,
and the topological transitions of fusion and fission
in the description of molecular sorting. This framework allowed us to explore this molecular distillation process as a self-organized phenomenon driven by mesoscopic physics.

The region of the parameter space where steady-state molecular sorting is viable is first identified. 
Higher pressure favors
large, round membranes. 
Lower induced curvature 
allows for the formation of large
vesicles, but is incompatible 
with sorting at higher pressures.
As a result
of these contrasting effects,
steady-state sorting can occur only within a restricted parameter region (Fig.~\ref{fig:heatmap_cprot_pr}).

By systematically varying membrane rigidity, 
spontaneous curvature, and pressure within the viable region, we examined how each parameter influences 
the speed of vesicle formation and  
the quality of the distillation process.
Larger curvature and rigidity accelerate vesicle formation, but reduce the time available for molecular reorganization, leading to lower sorting efficiency. Conversely, higher pressure favors sorting quality, but slows down 
sorting. These opposing effects establish a fundamental speed-quality trade-off.

To analyze this trade-off, we introduced the rate of negative entropy production as a criterion for optimal balance. Simulations reveal that this quantity exhibits a clear maximum at intermediate values of spontaneous curvature, pressure, and membrane rigidity, demonstrating the viability of the criterion. Within this framework, optimal parameters emerge naturally from the coupled dynamics of molecular aggregation and membrane mechanics.
It is interesting to speculate  
that living cells may have evolved to tune their mechanical and chemical 
properties to operate near 
maxima
of negative entropy production,
thus
achieving an efficient balance between rapid vesicle formation and high sorting fidelity.

We found that a characteristic vesicle size emerges dynamically from the interplay between pressure and curvature. 
Here, vesicle size is not set by an externally imposed threshold, but rather by competing forces: high pressure promotes larger vesicles, whereas high spontaneous curvature favors smaller ones. This underscores the self-organized character of membrane-driven molecular sorting, where vesicle formation arises from intrinsic system properties without requiring fine-tuned external regulation. 

Several questions remain open for future investigation. 
The one-dimensional implementation of the model 
here discussed captures 
important aspects of the coupling between molecular aggregation and membrane mechanics, while avoiding the substantial computational challenges associated with handling self-intersecting surfaces in higher dimensions. 
We expect the qualitative findings reported here, 
such as
the existence of a restricted sorting region, the speed-quality trade-off, and the emergence of an optimal balance maximizing negative entropy production, to remain valid in two dimensions. This expectation rests on a structural argument rather than on an explicit two-dimensional calculation. The free-energy functional in Eq.~\eqref{eq:G_GL} and the description of budding as thermally-activated escape over a curvature- and pressure-dependent barrier do not rely on the dimensionality of the membrane: the same ingredients, i.e.~a Landau-type functional coupling composition to curvature, a domain boundary tension controlling the activation barrier, and a Kramers-type extraction rate balanced against diffusive domain growth, reappear almost unchanged in a spherical-cap parametrization of a two-dimensional budding vesicle. What changes with dimensionality is mainly the geometry entering these ingredients, not the general mechanism, which rests on the same competition between line tension, spontaneous curvature and pressure along the budding pathway. A pressure threshold for vesicle formation also exists in higher dimensions~\cite{Hel73,MFB+91,FMR+94}, and the same qualitative dependence of the barrier on $c_0$, $\kappa$ and $p$ is also expected, although cumbersome calculations prevents from a closed analytical derivation. This structural correspondence is what leads us to expect the restricted sorting region, the speed-quality trade-off, and the entropy-production maximum identified here to persist, at least qualitatively, in two dimensions.\\
A natural perspective 
involves
the extension 
of  the present
investigation to  include
fully two-dimensional simulations. Such an extension would require explicitly modeling the lipid bilayer, the diffusing molecular species, and the curvature-generating mechanisms that drive vesicle formation, all while managing the topological complexities of fusion and fission. The present 
one-dimensional 
implementation of the model provides 
a simplified setting, in which parameter dependencies and identified governing principles 
can be explored
before tackling the computational challenges of more realistic two-dimensional simulations.\\
The experimental validation of the predicted speed-quality trade-off represents a crucial next step in bridging this theoretical framework with biological function. Our model identifies membrane rigidity, spontaneous curvature, and  pressure as  joint determinants of sorting efficiency.
These parameters are increasingly amenable to precise manipulation. For instance, membrane rigidity can be modulated through lipid composition or cholesterol content, spontaneous curvature by altering the properties of curvature-inducing proteins, and osmotic pressure by adjusting the ionic environment. Reconstituted systems, such as giant unilamellar vesicles with purified components, offer a controlled platform to test the predicted parameter regions, the
speed-quality trade-off, and the 
existence of a
proposed entropy-based optimum. The negative entropy production rate could be assessed by measuring the composition of incoming and outgoing vesicle populations, providing a quantitative bridge between theory and experiment. Such validation would not only test the framework but also offer insight into how cells tune their membrane properties to achieve efficient sorting.

\acknowledgments
Numerical calculations were made possible by a
SmartData@PoliTO agreement providing access to BIGDATA high-performing computing resources at Politecnico di
Torino. This research was funded
in whole, or in part, by the Austrian Science Fund (FWF)
grant I6533. For the purpose of open access, the author
has applied a CC BY public copyright licence to any
Author Accepted Manuscript version arising from this
submission. This work was supported by the Weave
project “Tissue material phase transitions and their role
in embryo pattern formation” from the Deutsche
Forschungsgemeinschaft (DFG, German Research
Foundation, 518354236, PE 3800/1-1) and \"Osterreichischer Wissenschaftsfonds (FWF, Austrian Science
Fund, I6533) (E. F).
\vskip -0.3cm

\appendix

\section{Numerical scheme}
\label{app:numerical_scheme}

In our numerical model, the membrane's energy is defined by Eq.~\ref{eq:discr_mem_ham}. The coupling between membrane curvature and the positions
of individual molecules
is established by the relation 
\begin{equation}\label{c_0k}
    c_{0,k} =  c_0^{\rm micro} \, \sigma_k \sigma_{k+1} \delta_{\sigma_k \sigma_{k+1}},
\end{equation}
so that
a spontaneous curvature~$c_0^{\rm micro}$ is 
associated with the presence of 
two molecules of the same species on adjacent
segments, mimicking the membrane bending effect induced by molecular crowding.

\begin{figure}[!b]
    \centering
    \includegraphics[width=\linewidth]{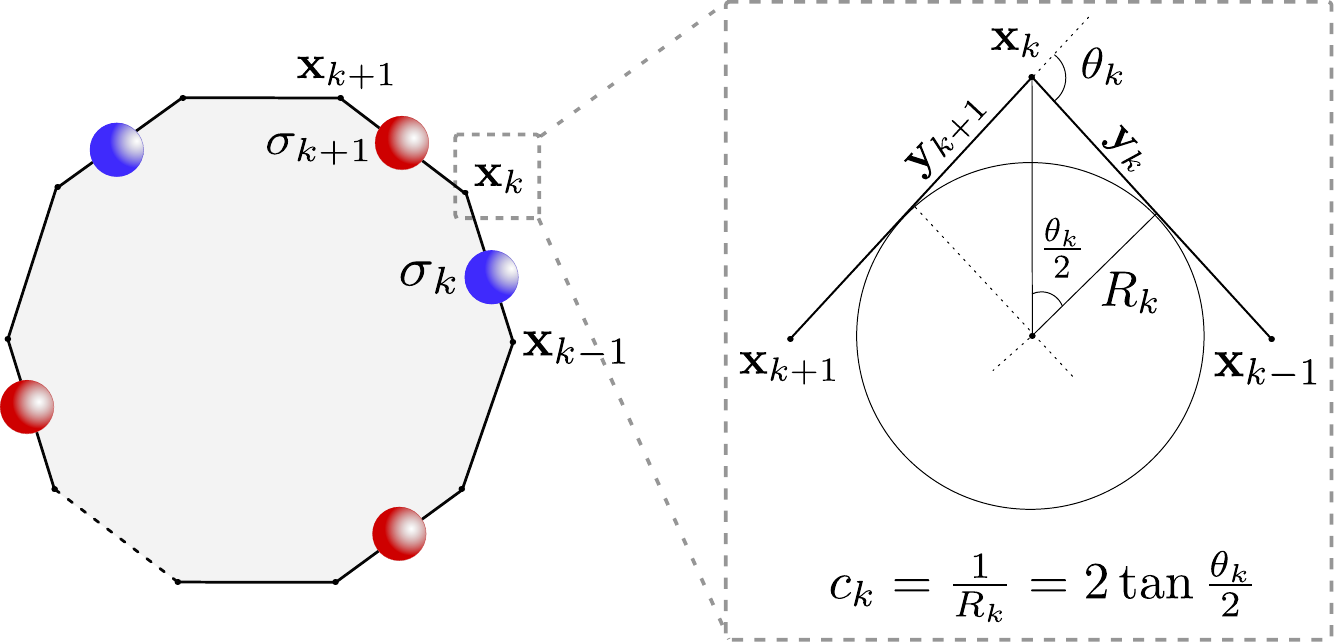}
    \caption{
    Schematic representation of the coupled membrane-molecule system at a given time $t$. The lipid membrane (solid black line) is modeled as a one-dimensional periodic lattice composed of $\mathbf{x}_k$ nodes connected by oriented edges $\mathbf{y}_k$, with $k \in \{1,2,...,N \}$, in the plane. The dashed black segment indicates that the lattice size $N$ changes when vesicle fission or vesicle fusion occur. The membrane is populated by a binary mixture of molecules $\sigma_k$ (shown as red and blue dots). In the inset, the osculating circle at node $k$ is shown, illustrating 
    the
    local curvature $c_k$. Each segment has unit length.}
    \label{fig:1dmodel}
\end{figure}

Let us consider the node~$\mathbf{x}_k$
located between consecutive oriented segments $\mathbf{y}_k$ and $\mathbf{y}_{k+1}$. The~discrete curvature~$c_k$ at node~$\mathbf{x}_k$ is defined as the reciprocal of the radius $R_k$ of the osculating circle shown in Fig.~\ref{fig:1dmodel}~\cite{BSS+00}. Using simple trigonometric relations and the constraint~$|\mathbf{y}_k|=a$ (with $a=1$), one finds
\begin{equation}
    c_k = 2 \tan \frac{\theta_k}{2},
    \label{eq:curvature}
\end{equation}
where $\theta_k$ is the angle shown in Fig.~\ref{fig:1dmodel}.

The area $A$ 
enclosed by the membrane 
is computed from the positions of the nodes using 
 textbook Gauss's formula for the area of a polygon.

\subsection{Dynamics of molecular inclusions}
Molecules diffuse and aggregate on the membrane via a discrete-space Markov process
modeling a
lattice-gas behavior. The occupation number $\sigma_k$ 
can be exchanged with its nearest neighbor $\sigma_{k+1}$, according to asymmetric Kawasaki dynamics~\cite{kaw66}. This exchange occurs at a rate
\begin{equation}
r=k_D e^{-\beta\Delta H_{\mathrm{mol}}},
\end{equation}
where $k_D$ is 
a diffusion rate, and $\Delta H_{\mathrm{mol}}$ is the energy difference associated with the exchange. Once the exchange rates are determined, the system evolves stochastically, with the timing of the next exchange event sampled using Gillespie’s algorithm~\cite{Gil76}.

\subsection{Membrane dynamics} 
Between two molecular jumps, the membrane system, composed of nodes $\mathbf x_k$ ($k=1,..., N$), evolves according to the overdamped Langevin equation
\begin{equation}
    \Gamma\mathbf{\dot x}_k(t)=- \nabla_{\mathbf{x}_k} H_{\mathrm{mem}}(t) +\sqrt{2 \Gamma k_BT}  \,\boldsymbol{\xi}_k(t),
    \label{eq:langevin}
\end{equation}
where the membrane energy is defined as in Eq.~\ref{eq:discr_mem_ham},~$\Gamma$ is a friction coefficient,~$k_B$ is the Boltzmann constant, and~$T$ is the temperature. 
To maintain a fixed (unit) distance between subsequent nodes $\mathbf x_{k-1}$ and $\mathbf x_k$, the stochastic force~$\boldsymbol \xi_k$, which represents thermal fluctuations, has the form $\boldsymbol \xi_k=\xi \, \mathbf k \times \mathbf{y}_k$, where $\mathbf k$ is the unit outward normal to the plane, $\mathbf y_k = \mathbf x_k -\mathbf x_{k-1}$, $|\mathbf{y}_k|=1$, and each component of $\boldsymbol{\xi}_k$
is an independent, zero-mean,
delta-correlated Gaussian noise of unit intensity.
The Langevin~equation (Eq.~\ref{eq:langevin}) governing  membrane dynamics is integrated 
numerically using the Heun-Stratonovich algorithm 
with an adaptive time step~\cite{BBT04,PTV+07}. For computational efficiency,
segments $\mathbf y_k$ in the plane are described as complex numbers \begin{equation}
    \omega_k=y_{k,1}+\mathrm{i} y_{k,2}= |\omega_k|e^{\mathrm{i}\phi_k}.
\end{equation}

In all simulations, the friction coefficient is set to $\Gamma=1$, 
which corresponds to measuring time in units of the system's intrinsic relaxation timescale. Thermal fluctuations are neglected by setting~$T=0$.

\subsection{Vesicle fusion}
Molecules are introduced into the system through random vesicle fusion events. Each incoming vesicle carries $n_A+n_B=7$ molecules, each assigned to one of the two species with equal probability. The number of empty sites in incoming vesicles is,
 on average, equal to the number of empty sites in budding vesicles, ensuring that the size of the lipid compartment remains stationary.  
To implement fusion, a lattice node is selected at random. It is then checked whether the
putative incoming vesicle, positioned tangentially to the compartment at that point, intersects with any other part of the membrane. If no intersection occurs, the insertion is accepted; otherwise, the procedure is repeated with another randomly selected node. An example of vesicle fusion is shown in Fig.~\ref{fig:example} (left box).

\subsection{Vesicle fission}
Molecules on the membrane form domains that induce curvature in the compartment, as illustrated in Fig.~\ref{fig:example}.
This curvature can lead to vesicle formation. When~a vesicle fully encloses, the membrane undergoes self-intersection. To detect such events, a~self-intersection check is performed at each integration step of membrane dynamics.
If an intersection is detected, the membrane is split at the intersection point, and the 
vesicle is removed from the simulation while
storing the number of extracted empty sites and of A and B molecules. The two open ends of the membrane compartment are then reconnected by introducing a new equidistant node, 
ensuring that the fixed length of membrane segments is preserved. 
After each fission event, the jump rates 
in the Gillespie algorithm are updated to reflect the new system configuration. An example of a vesicle fission event is shown in~Fig.~\ref{fig:example} (right box).

\subsection{Stationary state}\label{app:stationnary}
In each simulation, we discard the transient regime preceding attainment of the stationary state. To determine this, we fit the molecular density $\rho$ of the membrane compartment with the expression  
\(
\rho(t) = \rho_0 \left[ 1 - \exp\left(- {t}/{\tau} \right])\right]
\) 
and retain only times \( t > 2\tau \).

\section{Coarse-graining of the microscopic Hamiltonian}
\label{app:coarse_graining}

Here we derive the phenomenological functional $G_\mathrm{ph}$ of Eq.~\eqref{eq:1dcontinuous} from the microscopic lattice Hamiltonians $H_\mathrm{mem}$ and $H_\mathrm{int}$ of Eqs.~\eqref{eq:discr_mem_ham} and~\eqref{eq:discr_mol_ham}, and give explicit expressions for the phenomenological parameters $D,r,u,c_0$ in terms of the microscopic parameters $g$ (interaction strength), $\rho_0$ (mean molecular density) and $c_0^\mathrm{micro}$ (bare curvature induced by a pair of like neighbors).

\subsection{Coarse-grained fields}
Let $\ell$ be a coarse-graining length, in units of the lattice spacing $a$, with $1\ll\ell\ll N$. A coarse-graining cell $B_s$ centered at arc position $s$ contains $\ell$ consecutive edges. We define the coarse-grained molecular density, composition, and curvature fields
\begin{subequations}
\begin{align}
\rho(s) & =\frac{1}{\ell a}\sum_{k\in B_s}|\sigma_k|,\\
\psi(s) & =\frac{1}{\ell a\,\rho(s)}\sum_{k\in B_s}\sigma_k,\\
H(s) & =\frac{1}{\ell}\sum_{k\in B_s}c_k.
\end{align}
\end{subequations}
The coarse-grained free energy density $\mathcal{G}(s)$ is obtained by evaluating the energy of cell $B_s$ and dividing by its length $\ell a$.

\subsection{Coarse-graining the membrane term}
In the continuum limit $a\to0$, the bending term of $H_\mathrm{mem}$ becomes
\begin{equation}
\frac{\kappa}{2}\sum_k (c_k-c_{0,k})^2 a \;\longrightarrow\; \int_0^L \frac{\kappa}{2}\big[H(s)-c_0(s)\big]^2\,\mathrm{d}s,
\end{equation}
with $L=Na$. The local spontaneous curvature $c_{0,k}=c_0^\mathrm{micro}\,\sigma_k\sigma_{k+1}\delta_{\sigma_k,\sigma_{k+1}}$ is coarse-grained using the one-site marginals $p_A=\rho a(1+\psi)/2$, $p_B=\rho a(1-\psi)/2$ of the two species (independent-site approximation),
\begin{align}\nonumber
\langle c_{0,k}\rangle & \approx c_0^\mathrm{micro}\left(p_A^2+p_B^2\right)\\
& = c_0^\mathrm{micro}(\rho a)^2\,\frac{1+\psi^2}{2}
\equiv \bar c_0(\rho)\left(1+\psi^2\right),
\label{eq:c0_meanfield}
\end{align}
where $\bar c_0(\rho)=c_0^\mathrm{micro}(\rho a)^2/2$.  When the density is approximately stationary $\rho\approx \rho_0$,  the $\rho$-dependent, $\psi$-independent, part of $\bar c_0$ only renormalizes the curvature of the unsorted membrane and can be reabsorbed into a redefinition of $H$, writing $H \to H-\bar c_0$. Expanding for small $\psi$, the bending energy reduces to
\begin{equation}
 \frac{\kappa}{2} H^2 - \kappa\, H\,\bar c_0\,\psi^2 + O(\psi^4),
\label{eq:bending_expansion}
\end{equation}
which is the origin of the $\kappa c_0 H\psi^2$ coupling in Eq.~\eqref{eq:1dcontinuous}. The phenomenological spontaneous-curvature parameter appearing there can be assumed to be proportional to the coarse-grained term $\bar c_0(\rho_0)= c_0^\mathrm{micro}(\rho_0 a)^2/2$. The quartic term in \eqref{eq:bending_expansion} only renormalizes the quartic coefficient $u$ in the free-energy functional and it can be neglected.

\subsection{Coarse-graining the molecular term}
The energy term $H_\mathrm{int}$ describes a three-state (A, B, empty) lattice gas with same-species attraction. The corresponding mean-field free energy density per unit length, expressed in terms of $\rho$ and $\psi$, is
\begin{align}\nonumber 
\mathcal{G}_\mathrm{mol}(\rho,\psi) & = \frac{k_BT}{a}\left\{ (1-\rho a)\ln(1-\rho a) -\ln g\,(\rho a)^2\tfrac{1+\psi^2}{2} \right. \\
\nonumber & \quad + \rho a \ln\tfrac{\rho a}{2}\\
& \quad \left. + \rho a\Big[\tfrac{1+\psi}{2}\ln(1+\psi)+\tfrac{1-\psi}{2}\ln(1-\psi)\Big]
\right\},
\label{eq:fmol}
\end{align}
having used the coordination number $z=2$ appropriate to the 1D lattice. The expression of $\mathcal{G}_\mathrm{mol}$ can be expanded around the stationary mean density $\rho_0$ (which is not the free-energy minimum, but the self-organized density sustained by the non-equilibrium steady state) and around $\psi=0$, up to quartic order. After collecting the density-only, composition-only, gradient, and cross terms, this gives
\begin{align}\nonumber 
\mathcal{G}(\rho,\psi,H,\nabla\psi) & = \mu_\rho(\rho-\rho_0)+\frac{K_\rho}{2}(\rho-\rho_0)^2 \\
\nonumber & \quad + \rho_0\Big[\frac{\kappa_\psi}{2}|\nabla\psi|^2+\frac{r_0}{2}\psi^2+\frac{u_0}{4}\psi^4\Big] \\
& \quad + r_1(\rho-\rho_0)\psi^2 + \frac{\kappa}{2}\tilde H^2 -\kappa c_0\tilde H\psi^2,
\label{eq:full_f}
\end{align}
with
\begin{subequations}
\begin{align}
r_0 & =k_BT\rho_0\Big(1-\frac{\rho_0 a}{2}\ln g\Big), \label{eq:r0u0kphiA} \\
u_0 & =\frac{k_BT}{3}\rho_0, \label{eq:r0u0kphiB}\\
\kappa_\psi & =\frac{k_BT}{2}\rho_0 a^2\ln g,
\label{eq:r0u0kphiC}
\end{align}
\end{subequations}
while the specific expressions of the density chemical potential $\mu_\rho$, the density stiffness $K_\rho$, and the density-composition coupling $r_1$ obtained from the second derivatives of the density-only part of $f_\mathrm{mol}$ are not needed. Note that $\rho_0$ is not the equilibrium density: $\mu_\rho\neq0$ in general, reflecting the fact that $\rho_0$ is fixed self-consistently by the influx-outflux balance of the non-equilibrium steady state rather than by free-energy minimization.

\subsection{Effective Landau parameters}
Since density fluctuations relax fast compared to composition fluctuations, one can integrate out $\delta\rho=\rho-\rho_0$ at Gaussian order. This renormalizes the composition parameters,
\begin{equation}
D \equiv \rho_0\kappa_\psi,\quad
r \equiv \rho_0 r_0 -\frac{2\mu_\rho r_1}{K_\rho},\quad
u \equiv \rho_0 u_0 -\frac{2r_1^2}{K_\rho},
\label{eq:reff_ueff}
\end{equation}
which, together with $c_0 \propto \bar c_0(\rho_0)=c_0^\mathrm{micro}(\rho_0 a)^2/2$ from Eq.~\eqref{eq:c0_meanfield}, are exactly the parameters entering the phenomenological functional $G_\mathrm{ph}$ of Eq.~\eqref{eq:1dcontinuous}. Note from Eq.~\eqref{eq:r0u0kphiA} that $r_0$ changes sign at $\ln g = 2/(\rho_0 a)^2$, signaling the onset of the demixing transition within the present mean-field, independent-site, approximation. However, since the sorting process considered here is intrinsically out of equilibrium, molecular domains form dynamically through local aggregation even in parameter regions where an equilibrium demixing transition would not strictly occur. In this respect, we do not rely on the equilibrium mean-field prediction of a demixing transition anywhere else in this work.

\section{Vesicle budding}

\subsection{Budding condition}
\label{app:budding}
We first provide geometrical considerations and compute some quantities that will be useful later on. Let denote with $r$ the radius of the circular arc into which the domain segment is currently bent (use Fig.\ref{fig:fission_or_not} as a reference), and introduce the dimensionless coordinate $x=R_2/r$.
The arc spans a subtended angle $\theta(r)=2\pi x$, so $x =1$
corresponds to a fully closed circular vesicle of radius $R_2$.
By equating the expressions for the common chord one gets that the radius $R_M$ of the portion of the membrane without the domain satisfies
\begin{equation}
    R_M(x)\sin\left( \frac{\pi R_1}{R_M(x)} \right)=\frac{R_2}{x}\sin\left( \pi x \right).
\end{equation}
This radius goes from $R_1+R_2$ to $R_1$ during the budding process. In the limit $R_1\gg R_2$, we can get an explicit expression
\begin{equation}
    R_M(x)\approx R_1\left(1 +\frac{R_2}{R_1}\frac{1}{x\pi}\sin\left( \pi x\right)\right).
\end{equation}
The area enclosed by the intermediate arc (a circular segment) is 
\begin{equation}
A_\mathrm{arc}(x)=R_2^2\left[ \frac{\pi}{x}-\frac{\sin(2\pi x)}{2x^2}\right]
\end{equation}
and thus the total area enclosed by the membrane is 
\begin{align}
  \nonumber  A (x) & = \frac{R_M(x)^2}{2}\left[ \frac{2\pi R_1}{R_M(x)} - \sin{\left(\frac{2\pi R_1}{R_M(x)}\right)}\right] \\
    & \quad  + R_2^2\left[\frac{\pi}{x} - \frac{\sin{(2\pi x)}}{2x^2} \right]
\end{align}
that in the limit $R_1 \gg R_2$ reduces to 
\begin{equation}
A(x) \approx \pi R_1^2 + 2 R_1 R_2\frac{\sin{(\pi x)}}{x}  +R_2^2\left[ \frac{\pi}{x}-\frac{\sin(2\pi x)}{2x^2}\right].
\end{equation}

Using these expressions, we can write the membrane free energy as a function of $x$
\begin{equation}
    \begin{aligned}
        \label{eq:Gx}
        G(x)&=\pi\kappa R_{2}\left(\frac{x}{R_{2}}-c_{0}\right)^{2}+\pi\kappa R_{1}\left(\frac{1}{R_{M}\left(x\right)}\right)^{2}\\
        &\phantom{=} -p A(x)+2\gamma\left(H(x)\right)\Theta\left(1-x\right),
    \end{aligned}
\end{equation}
where the term containing the line tension $\gamma$ is generated by  like-molecule interactions and follows from the phenomenological approach of Sec.~\ref{sec:steady-state-condition}.
Notice that the Heaviside theta function removes the line tension once the vesicle has fully closed.

The free energy for initial configuration of Fig.~\ref{fig:fission_or_not} (lefthand) is
\begin{equation}
    \begin{aligned}
        G^{(i)}&=\kappa\pi R_{2}\left(\frac{1}{R_{1}+R_{2}}-c_{0}\right)^{2}+\kappa\pi R_{1}\left(\frac{1}{R_{1}+R_{2}}\right)^{2}\\
        &\phantom{=}- p\pi(R_{1}+R_{2})^{2}+2\gamma,
    \end{aligned}
\end{equation}
while for the final configuration (righthand of Fig.~\ref{fig:fission_or_not}) free energy is
\begin{equation}
    \begin{aligned}
        G^{(f)}&=\kappa\pi R_{2}\left(\frac{1}{R_{2}}-c_{0}\right)^{2}+\kappa\pi R_{1}\left(\frac{1}{R_{1}}\right)^{2}\\
        &\phantom{=}- p\left(\pi R_{1}^{2}+\pi R_{2}^{2}\right).
    \end{aligned}
\end{equation}
When $\Delta G=G^{(f)}-G^{(i)}<0$, the formation of the vesicle is energetically favored. This condition is satisfied if 
\begin{equation}
    p<p^*= \epsilon \left[ \frac{c_0 \kappa}{R_2^2} + \frac{\gamma}{\pi R_2^2} - \frac{\kappa}{2 R_2^3} \right] - \epsilon^2 \frac{c_0\kappa}{2R_2^2} + O\left(\epsilon^3\right)
    \end{equation}
    with $\epsilon = R_2/R_1$. Then assuming the natural expression of the radius of the formed vesicle $R_2 \sim c_0^{-1}$, we arrive to Eq.~\ref{eq:pressure_condition}.

\subsection{Size selection in the zero-temperature limit}
\label{app:activation_barrier}

The estimate above only compares the initial and final states, but the whole
shape of the free energy function $G(x,R_2)$ along the budding pathway
controls both whether budding occurs at all and, more finely, the rate at
which domains of different size are converted into vesicles.

At zero temperature, budding proceeds only along trajectories of
monotonically decreasing free energy: a domain of size $R_2$ can complete
fission if and only if $G(x,R_2)$ is monotonically decreasing over the whole
interval $[R_2/(R_1+R_2),1]$. For $p=0$, this condition holds whenever
$R_2>1/c_0$. For $p>0$, there exists a minimum value of $R_2$, larger than
$1/c_0$, above which the condition is satisfied. Within this admissible
range, all values of $R_2$ correspond to barrierless (purely downhill)
pathways; among them, we identify the typical vesicle size $R_2^*$ with the
value minimizing the free energy at closure, $G(1,R_2)$, which we expect to
dominate the outcome once the system is allowed to explore and settle into
the most stable accessible product. In the limit $R_1\gg R_2$, $R_2^*$ solves
the cubic equation
\begin{equation}
\kappa\left(c_0^2-\frac{1}{R_2^2}\right)-2pR_2=0.
\end{equation}
Computing the derivatives of $R_2$ on the parameters by the implicit function theorem one finds
\begin{subequations}
\begin{eqnarray}
  \frac{\partial R_2}{\partial p} & = & 
  \frac{R_2}{\frac{\kappa}{R_2^3} - p}\\
  \frac{\partial R_2}{\partial c} & = &  - \frac{c_0
  \kappa}{\frac{\kappa}{R^3} - p}\\
  \frac{\partial R_2}{\partial \kappa} & = &  - \frac{c_0^2 -
  \frac{1}{R_2^2}}{\frac{\kappa}{R_2^3} - p}
\end{eqnarray}
\end{subequations}
showing that for 
 $R_2>1/c_0$ (the necessary condition for budding) and
$p<\kappa/R_2^3$ (always verified in our simulations, see
also Eq.~\ref{eq:pressure_condition}),
$R_2$ decreases with $c_0$ and $\kappa$ (at non-zero~$p$) and increases with~$p$.
 
At finite temperature, budding becomes a thermally activated process occurring even
when $G(x,R_2)$ is not monotonic and controlled by the height of the
free-energy barrier $\Delta G^*(R_2)$, where the maximum of $G(x,R_2)$ can be at an
intermediate value of $x$.
Domains of different size then bud off at different rates, given by the Kramers
expression
\begin{equation}
J(R_2)\propto e^{-\beta\Delta G^*(R_2)}.
\label{eq:kramers}
\end{equation}
Note that $\Delta G^*(R_2)=0$ throughout the barrier-less region identified
above, so the process does not, by itself, select a unique size as
$\beta\to\infty$: within that region the zero-temperature selection is instead a
genuinely thermodynamic one, set by the depth of the final free-energy
minimum $G(1,R_2)$ rather than by the barrier.

\subsection{Sorting rate}
\label{app:sorting_rate}
The mean residence time of a molecule on the membrane splits into a diffusive contribution $\bar T_f$, spent freely diffusing before joining a domain, and a trapping contribution $\bar T_d$, spent waiting inside a growing domain for extraction.  A simple mean-first-passage-time calculation for diffusion gives 
the average time spent diffusing in the region of size $L$ between two absorbing domains. The mean time $\tau\left(s\right)$ for a molecule starting at distance $s$ from a domain to diffuse and hit domain boundary satisfies a Poisson equation $K_D\tau''\left(s\right)=-1$ subject to absorbing boundaries at the domain edges, $\tau\left(0\right)=\tau\left(L\right)=0$. Integrating twice and imposing the boundary conditions, we find $\tau\left(s\right)= s\left(L-s\right)/(2K_D)$. As the flux $\phi$  is uniform on the membrane, we have to integrate over the whole available interval, to obtain 
\begin{equation}
\bar{T}_{f}	 =\frac{1}{{L}}\int_{0}^{L}\tau\left(s\right)ds =\frac{L^{2}}{12K_D}.
\end{equation}

The average time $\bar{T}_{d}$ a molecule spends trapped inside a growing domain waiting for vesicle closure and bud off can be evaluated using the steady-state vesicle size distribution $P\left(R\right)$, 
\begin{subequations}
\begin{align}
\bar{T}_{d} & = \int_{0}^{\infty}\left[T_{{\rm growth}}\left(R\right)+T_{{\rm bud}}\left(R\right)\right]P\left(R\right)\mathrm{d}R\\
& = \int_{{0}}^{\infty}\left[\frac{R}{v}+\frac{1}{J\left(R\right)}\right]P\left(R\right)\mathrm{d}R
\end{align}
\end{subequations}
where $v$ is the constant speed of domain growth due to molecule diffusion on the membrane and absorption at the boundaries.
The time $\bar{T}_d$ grows with the size of the extracted vesicle. In particular, at very low temperature, when extracted vesicles have a well-defined typical size $R^*$, we can easily show that $\bar{T}_{d} \propto \frac{ R^* }{v}$. The first contribution $\frac{ R^* }{v}$ is trivial,  while the second  comes from noticing that
\begin{align}\nonumber 
\int_{{0}}^{\infty}\frac{1}{J\left(R\right)} P\left(R\right)\mathrm{d}R & = \frac{\int_{0}^{\infty}N_{\text{st}}(R)\,\mathrm{d}R}{\int_{0}^{\infty}J(R) N_{\text{st}}(R)\,\mathrm{d}R} \\
&  \approx \frac{1/L}{\phi/(2\pi \rho_d R^*)} \approx \frac{R^*}{v},
\end{align}
 where the growth speed is $v = \phi L /(2\pi \rho_d)$ by mass balance, under the same assumption that domains of constant density $\rho_d$ grow by molecule diffusion on the membrane.

\subsection{Sorting quality}
\label{app:quality}
Given the perimeter of the extracted vesicle, $2\pi R_{2}$, and the width  $\xi$ of domain walls, the average composition in the arc, defining the quality of the vesicle, is 
\begin{align}\nonumber 
q\left(x\right) & =\frac{1}{2\pi R_{2}}\left[\left(2\pi R_{2}-2\xi\right)\psi_{d}\left(x\right)\phantom{\int}\right.\\
& \left. \quad +2\int_{{\rm wall}}\psi_{d}\left(x\right)\tanh\left(\frac{s}{\xi}\right)\mathrm{d}s\right].
\end{align}
The wall contribution is of order $\xi/R_{2}$, providing a final result 
\begin{equation}
q\left(x\right)\approx\left(1-\frac{\xi\left(x\right)}{\pi R_{2}}\right)\psi_{d}\left(x\right).
\end{equation} 
Using $\psi_d\xi=\sqrt{2\rho_0\kappa_\psi/u}$ (independent of $R_2$ throughout the budding pathway) and $\psi^*=\sqrt{(\kappa c_0/R_1-r)/u}$ for the composition reached deep inside a large, slowly-closing domain, in the limit $R_2\ll R_1$, we get 
\begin{equation}
q(R_2) \approx  \psi^* - \frac{1}{\pi R_2}\sqrt{\frac{2\rho_0\kappa_\psi}{u}}.
\label{eq:quality_R}
\end{equation}
Equation~\eqref{eq:quality_R} shows that the quality of an extracted vesicle increases with its size: larger, slower-forming domains have more time for the two molecular species to demix before fission occurs. 


%

\end{document}